\documentclass[twocolumn]{aastex7}
\usepackage{bm}
\usepackage{braket}
\usepackage{amsmath}
\usepackage{subeqnarray}
\usepackage{here}
\usepackage{multirow}
\usepackage{makecell}
\usepackage{tabularx}
\usepackage{threeparttable}
\usepackage{lipsum}
\usepackage{hyperref}
\hypersetup{colorlinks=true,linkcolor=blue,citecolor=blue,filecolor=blue,urlcolor=blue,}
\usepackage{soul}
\usepackage{tabularx}

\submitjournal{ApJ}
\shorttitle{}
\shortauthors{Suzuki et al.}
\begin{document}
\title{Influence of Alfv\'{e}n wave turbulence dissipation in stellar winds of the Sun and M dwarfs}

\correspondingauthor{Takaaki Yokoyama}
\email{yokoyama.takaaki.2a@kyoto-u.ac.jp}
\author{Kaito Suzuki}
\affiliation{\rm{Department of Astronomy, Kyoto University, Sakyo-ku, Kyoto 606-8502, Japan}}
\email{kaito@kusastro.kyoto-u.ac.jp}

\author[0000-0001-5457-4999]{Takaaki Yokoyama}
\affiliation{\rm{Astronomical Observatory, Kyoto University, Sakyo-ku, Kyoto 606-8502, Japan}}
\email{yokoyama.takaaki.2a@kyoto-u.ac.jp}

\author[0000-0002-7136-8190]{Munehito Shoda}
\affiliation{\rm{Department of Earth and Planetary Science, School of Science, The University of Tokyo, Hongo, Bunkyo-ku, Tokyo, 113-0033, Japan}}
\email{shoda.m.astroph@gmail.com}

\begin{abstract}
The physics of M-dwarf stellar winds is increasingly being investigated within the framework of wave-driven models. In the case of the solar wind, Alfv\'en-wave turbulence (AWT) is believed to play a crucial role in heating and accelerating the plasma. In this study, we perform one-dimensional magnetohydrodynamic simulations of the solar and M-dwarf winds using a model that incorporates shock heating generated by mode conversion and treats AWT as an optional component. For the solar wind, we confirm that including AWT reproduces the observed predominance of outwardly propagating Alfv\'en waves, supporting the role of turbulent dissipation in driving the wind. We further validate this AWT-driven solar wind directly against \textit{Parker Solar Probe} observations, finding good agreement in both the location of the Alfv\'en point and the level of the normalized cross helicity in the fast solar wind. In M-dwarf simulations, we find that AWT dominates the overall heating budget within the investigated parameter range. A parameter survey of coronal magnetic field strength reveals that the dependence of the mass-loss rate on $B_{\mathrm{cor}}$ differs qualitatively between cases with and without AWT, highlighting the importance of chromospheric dissipation. In the absence of AWT, the resulting mass-loss rates are broadly consistent with previous studies and observations. In contrast, when AWT is included, the mass-loss rate is significantly reduced because turbulent dissipation decreases the Alfv\'en-wave (Poynting) flux entering the corona. These results suggest that the effects of chromospheric turbulence play a crucial role in modeling M-dwarf winds.
\end{abstract}

\keywords{magnetohydrodynamics(MHD) - stellar winds - solar winds - Alfv\'{e}n waves}

\section{Introduction} \label{sec:intro}
Stellar winds are plasma outflows from stars that extend into interplanetary space \citep{Parker1958, Lamers_1999_book}. They play crucial roles in the evolution of stars and planetary systems in multiple ways. For instance, magnetized winds transport stellar angular momentum outward \citep{Weber_1967_ApJ, Sakurai_1985_AandA, Reville_2015_ApJ}, leading to the stellar spindown \citep{1972ApJ...171..565S, Barnes_2003_ApJ, 2016Natur.529..181V}. They also shape planetary evolution by influencing atmospheric loss, with both positive and negative effects \citep{Bourrier2016, Vidotto2017}. Stellar winds are therefore essential factors in describing the evolution of star--planet systems.

One of the key parameters characterizing stellar winds is the mass-loss rate. Although stellar wind mass-loss rates have been indirectly measured for stars with various activity levels \citep{Wood_2002_ApJ, Wood_2005_ApJ, Jardine_2019_MNRAS, Wood_2021_ApJ, Kislyakova_2024_NatAs}, the number of samples remains limited and the uncertainties are large. Consequently, theoretical studies are regarded as essential alongside observations.

Among stellar wind models, the most successful is the so-called Alfv\'en-wave model. In this scenario, Alfv\'en waves generated by convective motions at the photosphere propagate along open magnetic field lines into the solar atmosphere and dissipate their energy \citep{Hollwerg_1986_JGR}. While the exact mechanisms of Alfv\'en-wave dissipation remain debated \citep{van_Ballegooijen_2016_ApJ, Magyar_2017_NatSR, Shoda_2019_ApJ}, the Alfv\'en-wave model is known to reproduce (at least part of) the solar wind \citep{suzuki2005making, van_der_Holst_2014_ApJ}. Nevertheless, alternative energy sources, distinct from convection-driven Alfv\'en waves, have also gained attention in recent years \citep{Wang_2020_ApJ, Bale_2023_Natur, Iijima_2023_ApJ}.

A growing number of studies have explored Alfv\'en-wave--driven models to estimate the wind mass-loss rates of stars beyond the Sun \citep{Mesquita_2020_MNRAS, Shoda2020, Evensberget_2021_MNRAS, Reville_2024_ApJ}. For example, \cite{Suzuki_2013_PASJ} derived a scaling law for the mass-loss rate of the young Sun using one-dimensional MHD simulations. Building on this line of work, \cite{Sakaue2021, 2021ApJ...906L..13S} extended the solar wind model developed by \cite{Sakaue2020} and performed one-dimensional MHD simulations of stellar winds for M dwarfs. However, Alfv\'en-wave turbulence (AWT) was not included in their model; instead, the winds were driven by slow shocks generated through mode conversion from Alfv\'en waves \citep{suzuki2005making}. There has been a study that derived a mass-loss rate scaling while incorporating AWT \citep{Cranmer_2011_ApJ}, but it relied on phenomenological treatments of wave propagation, such as the WKB approximation in the chromosphere, and its validity in this regime may be subject to some uncertainty. Therefore, performing numerical simulations that explicitly incorporate AWT and chromospheric wave propagation will be essential for advancing our understanding of stellar-wind dynamics in M dwarfs. In a concurrent study, \citet{Shoda_2026_MNRAS} tested the effect of suppressing chromospheric turbulent dissipation using a broad ensemble of solar magnetic flux tubes spanning a wide range of coronal field strengths, used as a proxy for varying stellar magnetic activity, and found that suppressing this dissipation brings the resulting wind mass flux into better agreement with the empirically observed relation between magnetic field strength and mass flux.

In this study, we use a one-dimensional MHD model that self-consistently includes shock heating generated by mode conversion and in which the contribution of AWT can optionally be included or excluded. This framework allows us to investigate the relative roles of shock heating and AWT in heating and accelerating the solar wind and the winds of M dwarfs. For the solar wind, we conduct two sets of simulations: one including AWT throughout the entire domain from the photosphere to the heliosphere, and one excluding AWT. For M-dwarf winds, we likewise perform simulations with and without AWT to examine whether turbulent dissipation is the dominant heating mechanism. In our model, AWT is active not only in the corona but also in the chromosphere, in contrast to \citet{shoda2018self}, where chromospheric dissipation was neglected. Furthermore, to clarify how AWT affects global wind properties, we carry out a parameter survey varying the coronal magnetic field strength and compare the resulting mass-loss rates between the cases with and without AWT.

The remainder of this paper is organized as follows. Section~\ref{chap_model} describes the numerical settings. In Section~\ref{sec:results}, we first validate our model using solar-wind simulations with and without AWT and then present the results of applying the model to M-dwarf winds. In Section~\ref{sec:Discussion}, we discuss quantitative comparisons with previous studies, interpret the numerical results, and highlight the implications of our findings.

\section{Simulation model}\label{chap_model}

In this study, we select the Sun and an M-dwarf GJ 436 as the simulation targets (Table~\ref{tb_parameter}). The effect of stellar rotation is neglected, as both stars are slow rotators \citep{Sakurai1985, Bourrier2018}. The surface (photospheric) quantities, indicated by the subscript ``ph'', are used as control parameters.

The surface temperature $T_\mathrm{ph}$ is set equal to the effective temperature $T_\mathrm{eff}$. Since the photosphere is defined as the layer where the optical depth equals unity, the surface mass density $\rho_\mathrm{ph}$ is estimated using the optical depth equation \citep{Kippenhahn_1990_book, Cranmer_2011_ApJ}
\begin{align}
    \kappa_{\rm R} \left( \rho_{\rm ph}, T_{\rm ph} \right) \rho_{\rm ph} H_{\rm ph} = \frac{2}{3},
\end{align}
where $H_{\rm ph}$ is the pressure scale height on the photosphere and $\kappa_{\rm R}$ is the Rosseland opacity. We use the Rosseland opacity table from \citet{Shoda_2025_AandA}, assuming the stellar metallicity equals the solar value. The surface pressure $p_\mathrm{ph}$ is derived from the equation of state for neutral hydrogen gas:
\begin{align}
    p_{\rm ph} = \rho_{\rm ph} k_{\rm B} T_{\rm ph} / m_{\rm H},
\end{align}
where $k_{\rm B}$ is Boltzmann constant and $m_{\rm H}$ is the mass of hydrogen. We adopt the equation of state for neutral hydrogen here because, at the photospheric temperatures considered ($T_\mathrm{ph}=5770$\,K for the Sun and $3416$\,K for the M dwarf), the hydrogen ionization fraction is negligible, so the mean molecular weight is close to unity. This calculation is used solely to obtain realistic values of the mass density $\rho_\mathrm{ph}$ and pressure $p_\mathrm{ph}$ at the inner boundary; the single-fluid MHD equations [Equations~(\ref{eq_basic1})--(\ref{eq_basic7})], into which these values enter as boundary conditions, depend only on $\rho$ and $p$ as state variables and do not themselves reference the ionization state. The ionization of hydrogen, which becomes significant above the photosphere, is instead incorporated self-consistently throughout the computational domain through the equation of state $e^{\mathrm{int}}(p,\rho)$, which accounts for the effect of ionization \citep{Voegler2004, shoda2021corona}. The surface magnetic field $B_{r,\mathrm{ph}}$ is assumed to be in pressure balance with the surrounding gas, i.e., its magnetic pressure equals $p_\mathrm{ph}$. The transverse correlation length of turbulence at the surface is taken proportional to the pressure scale height \citep{Cranmer_2011_ApJ} and normalized to $\lambda_{\perp,\mathrm{ph}} = 150$ km for the Sun, giving $\lambda_{\perp,\mathrm{ph}} = 36$ km for GJ 436.
\begin{deluxetable}{lcc}
\setlength{\tabcolsep}{16pt}
\tablecaption{Parameters of the Sun and M-dwarf \label{tb_parameter}}
\tablehead{\colhead{} & \colhead{Sun} & \colhead{M-dwarf}}
\startdata
$M_*[M_{\odot}]$ & 1.0 & 0.507 \\
$r_\ast[r_{\odot}]$ & 1.0 & 0.455 \\
$T_{\mathrm{ph}}\,[\mathrm{K}]$ & 5770 & 3416 \\
$\rho_\mathrm{ph}\,[\mathrm{g\ cm}^{-3}]$ & $1.87\times 10^{-7}$ & $2.73\times 10^{-5}$ \\
$B_{r,\mathrm{ph}}\,[\mathrm{G}]$ & $1.34\times 10^3$ & $1.39\times 10^4$ \\
$H_{\mathrm{ph}}\,[\mathrm{km}]$ & 174 & 42 \\
$\lambda_{\perp,\mathrm{ph}}\,[\mathrm{km}]$ & 150 & 36 \\
\enddata
\tablecomments{
Stellar mass $M_\ast$ and radius $r_\ast$ are given in solar units (subscript $\odot$; \citet{Vidotto2017}). The photospheric temperature is set equal to the effective temperature $T_\mathrm{eff}$ \citep{Braun2012}.}
\end{deluxetable}

We conducted two runs (Table \ref{tb_parameter2}) for solar winds and
14 runs (Table \ref{tb_parameter3}) for M-dwarf winds including
parameter survey on coronal magnetic field strength $B_{\mathrm{cor}}$.

\subsection{Numerical domain and basic equations}\label{sec_model1}

Numerical simulations are based on the previous study by \citet{Shoda2023}. We consider a one-dimensional single open flux tube extending radially from the stellar photosphere. The coordinate $r$ denotes the tube axis originating at the stellar center, while $x$ and $y$ denote the perpendicular directions, forming a right-handed orthogonal coordinate system $(x, y, r)$. The numerical domain extends from the stellar surface to a location well beyond the Alfv\'en point of each solution. The basic equations are given as follows:
\begin{align}
\label{eq_basic1}
	\frac{\partial}{\partial t} \rho + \frac{1}{r^{2}f^{\mathrm{open}}} \frac{\partial}{\partial r}(\rho v_{r} r^{2} f^{\mathrm{open}}) = 0,
\end{align}
\begin{align}
\label{eq_basic2}
	\frac{\partial}{\partial t} \rho v_{r} + 
	\frac{1}{r^{2}f^{\mathrm{open}}} \frac{\partial}{\partial r} \left[ (\rho v_{r}^{2} + p_{\mathrm{T}}) r^{2} f^{\mathrm{open}} \right] \notag \\
    = \left( p + \frac{1}{2}\rho v_{\perp}^{2}\right) \frac{d}{dr} \ln(r^{2}f^{\mathrm{open}}) - \rho \frac{GM_{*}}{r^{2}},
\end{align}
\begin{align}
\label{eq_basic3}
	\frac{\partial}{\partial t} \rho v_{x} + 
	\frac{1}{r^{2}f^{\mathrm{open}}} \frac{\partial}{\partial r} \left[ \left( \rho v_{r} v_{x} - \frac{B_{r}B_{x}}{4\pi}\right) r^{2} f^{\mathrm{open}} \right] \notag \\ 
    = \frac{1}{2}\left( -\rho v_{r}v_{x} + \frac{B_{r}B_{x}}{4\pi}\right) \frac{d}{dr} \ln(r^{2}f^{\mathrm{open}}) + \rho D_{x}^{\mathrm{v}},
\end{align}
\begin{align}
\label{eq_basic4}
	\frac{\partial}{\partial t} \rho v_{y } + 
	\frac{1}{r^{2}f^{\mathrm{open}}} \frac{\partial}{\partial r} \left[ \left( \rho v_{r} v_{y} - \frac{B_{r}B_{y}}{4\pi}\right) r^{2} f^{\mathrm{open}}\right] \notag \\ 
    = \frac{1}{2}\left( -\rho v_{r}v_{y} + \frac{B_{r}B_{y}}{4\pi}\right) \frac{d}{dr} \ln(r^{2}f^{\mathrm{open}}) + \rho D_{y}^{\mathrm{v}},
\end{align}
\begin{align}
\label{eq_basic5}
	\frac{\partial}{\partial t} B_{x}  + 
	\frac{1}{r^{2}f^{\mathrm{open}}} \frac{\partial}{\partial r}\left[ (v_{r}B_{x} - v_{x}B_{r}) r^{2} f^{\mathrm{open}}\right] \notag \\ 
    = \frac{1}{2}(v_{r}B_{x} - v_{x}B_{r})\frac{d}{dr} \ln(r^{2}f^{\mathrm{open}}) + \sqrt{4\pi \rho} D_{x}^{\mathrm{b}},
\end{align}
\begin{align}
\label{eq_basic6}
	\frac{\partial}{\partial t} B_{y} + 
	\frac{1}{r^{2}f^{\mathrm{open}}} \frac{\partial}{\partial r}\left[ (v_{r}B_{y} - v_{y}B_{r}) r^{2} f^{\mathrm{open}}\right] \notag \\
    = \frac{1}{2}(v_{r}B_{y} - v_{y}B_{r})\frac{d}{dr} \ln(r^{2}f^{\mathrm{open}}) + \sqrt{4\pi \rho} D_{y}^{\mathrm{b}},
\end{align}
\begin{align}
\label{eq_basic7}
	\frac{\partial}{\partial t} e + 
	\frac{1}{r^{2}f^{\mathrm{open}}} \frac{\partial}{\partial r}\left[ \left\{ (e + p_\mathrm{T})v_{r} - B_{r} \frac{\bm{v}_{\perp} \cdot \bm{B}_{\perp}}{4\pi}\right\} r^{2} f^{\mathrm{open}}\right] \notag \\ 
    = - \rho v_{r} \frac{GM_{*}}{r^{2}} + Q^{\mathrm{cnd}} + Q^{\mathrm{rad}},
\end{align}
where $\rho$ denotes the mass density, $p$ the gas pressure, $v_r$ and $B_r$ the radial components of velocity and magnetic field, respectively, and $\bm{v}_\perp$ and $\bm{B}_\perp$ the perpendicular components defined as
\begin{equation}
\label{eq_basic8}
	\bm{v}_{\perp} = v_{x}\bm{e}_{x} + v_{y}\bm{e}_{y},\quad \bm{B}_{\perp} = B_{x}\bm{e}_{x} + B_{y}\bm{e}_{y},
\end{equation}
where $\bm{e}_{x}$ and $\bm{e}_{y}$ are the unit vectors in $x$ and $y$, respectively.
The total pressure $p_{\mathrm{T}}$ and total energy density $e$ are expressed as
\begin{equation}
\label{eq_basic9}
	p_{\mathrm{T}} = p + \frac{\bm{B}_{\perp}^{2}}{8\pi},\quad e = e^{\mathrm{int}} + \frac{1}{2}\rho \bm{v}^{2} + \frac{\bm{B}_{\perp}^{2}}{8\pi},
\end{equation}
where $e^{\mathrm{int}}$ is the internal energy density per unit volume and is determined by the equation of state including the effect of ionization as a function of $p$ and $\rho$ \citep{Voegler2004, shoda2021corona}.

$f^{\mathrm{open}}$ represents the filling factor of the open magnetic flux, as defined in Section \ref{sec_model3}. The terms $D_{i}^{\mathrm{v}}$ and $D_{i}^{\mathrm{b}}$ ($i=x$ or $y$) correspond to the phenomenological AWT terms introduced in Section \ref{sec_model2}. The heating rates by thermal conduction and radiation are expressed as $Q^{\mathrm{cnd}}$ and $Q^{\mathrm{rad}}$, respectively, in Section \ref{sec_model heating}. $G$ denotes the gravitational constant.

\subsection{Flux tube}\label{sec_model3}

The filling factor of the flux tube, $f^{\mathrm{open}}$, is prescribed as a function of $r$. As the flux tube expands laterally with increasing $r$, the filling factor also increases. In other words, the filling factor determines the degree of superradial expansion of the flux tube with radial distance. We express $f^{\mathrm{open}}$ as a function of $r$ as follows:
\begin{equation}
\label{eq_ff1}
	f^{\mathrm{open}}(r) = f_{\mathrm{ph}}^{\mathrm{open}} \frac{f_{1}(r)f_{2}(r)}{\sqrt{f_{1}(r)^{2} + f_{2}(r)^{2}}},
\end{equation}
\begin{equation}
\label{eq_ff2}
	f_{1}(r) = \frac{\eta ^{\mathrm{exp}}}{\sqrt{(\eta ^{\mathrm{exp}})^{2} - 1}} \exp\left(\frac{r-r^\mathrm{expch}}{H^{\mathrm{expch}}}\right),
\end{equation}
\begin{equation}
\label{eq_ff3}
	f_{2}(r) = \frac{\mathcal{F}(r) + \eta ^{\mathrm{exp}}f_{\mathrm{ph}}^{\mathrm{open}} + \mathcal{F}(r_{*})\left(\eta ^{\mathrm{exp}}f_{\mathrm{ph}}^{\mathrm{open}} - 1\right)}{f_{\mathrm{ph}}^{\mathrm{open}}\left(\mathcal{F}(r) + 1\right)},
\end{equation}
where
\begin{equation}
\label{eq_ff4}
	\mathcal{F}(r) = \exp\left(\frac{r - r^\mathrm{expcr}}{H^\mathrm{expcr}}\right).
\end{equation}
We note that $f(r)$ is normalized so that $\lim_{r \to \infty} f(r) = 1$. The functions $f_1(r)$ and $f_2(r)$ describe the flux-tube expansion in the chromosphere \citep{Ishikawa2021} and in the lower corona \citep{Kopp1976}, respectively. For the chromospheric expansion, we set $r^\mathrm{expch}=r_\ast$ and $H^\mathrm{expch}=2 H_\mathrm{ph}$, while for the coronal expansion they are $r^\mathrm{expcr}=1.3 r_\ast$ and $H^\mathrm{expcr}=0.5 r_\ast$. The surface value $f_{\mathrm{ph}}^{\mathrm{open}}$ and the chromospheric expansion parameter $\eta ^{\mathrm{exp}}$ are determined indirectly by specifying the coronal magnetic field strength $B_{\mathrm{cor}}$ and the ratio of open to surface magnetic flux $\xi^{\mathrm{open}}$.
\begin{equation}
\label{eq_ff5}
	f_{\mathrm{ph}}^{\mathrm{open}} = \xi^{\mathrm{open}}\frac{B_{\mathrm{cor}}}{B_{r,\mathrm{ph}}},\quad \eta^{\mathrm{exp}} = \frac{B_{r,\mathrm{ph}}}{B_{\mathrm{cor}}},
\end{equation} 
where $\xi^{\mathrm{open}} = 0.1$ is assumed. The radial component of the magnetic field is deduced from $f^{\mathrm{open}}(r)$ from the magnetic-flux conservation:
\begin{equation}
\label{eq_br}
	B_{r}(r) = B_{r,\mathrm{ph}}\frac{r_{*}^{2}f_{\mathrm{ph}}^{\mathrm{open}}}{r^{2}f^{\mathrm{open}}(r)}.
\end{equation}

\subsection{Alfv\'{e}n-wave turbulence dissipation}\label{sec_model2}

Following the approach of \cite{shoda2018self}, but with a slightly modified formulation, 
the AWT dissipation terms $D_{i}^\mathrm{v}$ and $D_{i}^\mathrm{b}$ are expressed phenomenologically 
in terms of Els\"asser variables \citep{elsasser1950hydromagnetic} as
\begin{equation}
    \label{eq_turb1}
    D_{i}^{\mathrm{v}} = -\frac{c_{\mathrm{d}}}{4\lambda_{\perp}}\bigl(|z_\perp^{+}|z_{i}^{-} + |z_\perp^{-}|z_{i}^{+}\bigr),
\end{equation}
\begin{equation}
    \label{eq_turb2}
    D_{i}^{\mathrm{b}} = -\frac{c_{\mathrm{d}}}{4\lambda_{\perp}}\bigl(|z_\perp^{+}|z_{i}^{-} - |z_\perp^{-}|z_{i}^{+}\bigr),
\end{equation}
where $i \in \{x,y\}$, and
\begin{equation}
    \label{eq_turb3}
    \bm{z}_{\perp}^{\pm} = \bm{v}_{\perp} \mp \frac{\bm{B}_{\perp}}{\sqrt{4\pi\rho}}
= z_x^{\pm} \bm{e}_x + z_y^{\pm} \bm{e}_y,
\end{equation}
where $\bm{z}_\perp^{+}$ ($\bm{z}_\perp^{-}$) denote the outward (inward) Els\"asser variables.
The dimensionless coefficient is set to $c_{\mathrm{d}}=0.1$ \citep{Ballegooijen2016}.
The perpendicular correlation length $\lambda_{\perp}$ is assumed to scale with the flux-tube radius as
\begin{equation}
    \label{eq_lambda}
    \lambda_{\perp}(r) = \lambda_{\perp,\mathrm{ph}} \sqrt{\frac{B_{r,\mathrm{ph}}}{B_{r}(r)}}.
\end{equation}

\subsection{Heating terms}\label{sec_model heating}

The heating due to thermal conduction is given by
\begin{equation}
    \label{eq_cnd1}
    Q^{\mathrm{cnd}} = -\frac{1}{r^{2}f^{\mathrm{open}}} 
\frac{\partial}{\partial r}\!\left(F^{\mathrm{cnd}} r^{2}f^{\mathrm{open}}\right),
\end{equation}
where $F^{\mathrm{cnd}}$ denotes the heat flux. 
To compute $F^{\mathrm{cnd}}$, we use the Spitzer--H\"arm form of the conductive flux 
\citep{1953PhRv...89..977S}, with quenching applied in low-density regions \citep{Shoda2020}:
\begin{equation}
    \label{eq_cnd2}
    F^{\mathrm{cnd}} = -\kappa_{\mathrm{SH}}T^{5/2} 
\frac{\partial T}{\partial r} \min\!\left[1, \left( \frac{r}{r^{\mathrm{cnd}}} \right)^{-2} \right],
\end{equation}
where $\kappa_{\mathrm{SH}} = 10^{-6}\ \mathrm{erg\ cm^{-1}\ s^{-1}\ K^{-7/2}}$ 
and $r^{\mathrm{cnd}}$ denotes the quenching location 
(see Tables \ref{tb_parameter2} and \ref{tb_parameter3}).

The radiative heating term $Q^{\mathrm{rad}}$ consists of two components: 
an optically thick contribution $Q^{\mathrm{thick}}$ and an optically thin contribution $Q^{\mathrm{thin}}$, 
combined as \citep{shoda2021corona}
\begin{equation}
    \label{eq_rad1}
    Q^{\mathrm{rad}} = (1-\xi^{\mathrm{rad}})Q^{\mathrm{thick}} + \xi^{\mathrm{rad}}Q^{\mathrm{thin}},
\end{equation}
\begin{equation}
    \label{eq_rad2}
    \xi^{\mathrm{rad}} = \min\!\left[ 1, \exp\!\left( -\frac{p}{p^{\mathrm{rad}}}\right)\right],
\end{equation}
where $p^{\mathrm{rad}}/p_{\mathrm{ph}}$ is set to 0.1.
The optically thick term $Q^{\mathrm{thick}}$ relaxes the internal energy toward a reference value over a specified timescale 
and is approximated as \citep{gudiksen2005abheating}
\begin{equation}
    \label{eq_rad3}
    Q^{\mathrm{thick}} = -\frac{1}{\tau^{\mathrm{thick}}}(e^{\mathrm{int}} - e^{\mathrm{int}}_{\mathrm{ref}}),
\end{equation}
where the timescale $\tau^{\mathrm{thick}}$ is given by
\begin{equation}
    \label{eq_rad4}
    \tau^{\mathrm{thick}} = \tau^{\mathrm{thick}}_0 \left(\frac{\rho}{\rho_\mathrm{ph}}\right)^{-1/2}.
\end{equation}
Here, $\tau^{\mathrm{thick}}_0$ is set to $0.1\ \mathrm{s}$, 
and $e^{\mathrm{int}}_{\mathrm{ref}}$ denotes the reference internal energy corresponding to the photospheric temperature $T_\mathrm{ph}$.
Thus, the optically thick cooling term relaxes the atmospheric temperature toward the surface value.

The optically thin cooling rate $Q^{\mathrm{thin}}$ is given in terms of the radiative loss function $\Lambda(T)$ as
\begin{equation}
    \label{eq_rad5}
    Q^{\mathrm{thin}} = -n_\mathrm{H} n_\mathrm{e} \Lambda(T),
\end{equation}
where $n_\mathrm{H}$ and $n_\mathrm{e}$ are the number densities of hydrogen atoms and electrons, respectively, computed from the equation of state. The radiative loss function $\Lambda(T)$ is calculated using version 10 of the CHIANTI atomic database \citep{Dere_1997_AAS, Del_Zanna_2021_ApJ} with photospheric abundances, and is extended to the lower-temperature regime ($T<10^{4}\ \mathrm{K}$) following \cite{goodman2012radiating}, as described in \cite{shoda2021corona}.

Our model accounts for the heating due to AWT, even though it does not appear explicitly in the energy equation 
(\ref{eq_basic7}). This can be seen by writing the internal energy equation as
\begin{align}
    &\frac{\partial e^\mathrm{int}}{\partial t} +
    \frac{1}{r^{2}f^{\mathrm{open}}} \frac{\partial}{\partial r}\!\left(e^\mathrm{int} v_r r^{2}f^{\mathrm{open}} \right) \notag \\
    &= -\frac{p}{r^{2}f^{\mathrm{open}}}\frac{\partial}{\partial r} \left( v_r r^{2}f^{\mathrm{open}} \right)
    + Q^{\mathrm{turb}} + Q^{\mathrm{cnd}} + Q^{\mathrm{rad}},
\end{align}
where the turbulent heating rate is given by
\begin{align}
\label{eq_Qturb}
    Q^{\mathrm{turb}}
    &= -\rho \!\left( v_{x}D_{x}^\mathrm{v} + v_{y}D_{y}^\mathrm{v}
    + \frac{B_{x}}{\sqrt{4\pi\rho}}D_{x}^\mathrm{b} 
    + \frac{B_{y}}{\sqrt{4\pi\rho}}D_{y}^\mathrm{b} \right) \notag \\
    &= -\frac{c_{\rm d}}{4 \lambda_\perp}
    \left( \left| z_\perp^+ \right| {z_\perp^-}^{2}
    + \left| z_\perp^- \right| {z_\perp^+}^{2} \right).
\end{align}
Thus, by solving the basic MHD equations, the heating due to AWT is naturally included. 
The same argument applies to numerical dissipation and heating, which implicitly account for the effects of shock waves.

\subsection{Boundary conditions}\label{sec_model4}

A free boundary condition is imposed at the outer boundary $r = r_\mathrm{out}$. 
At the inner boundary, located at the stellar surface $r_\mathrm{in}=r_\ast$, fixed values are specified for $\rho_\mathrm{in}=\rho_\mathrm{ph}$, $p_\mathrm{in}=p_\mathrm{ph}$, and $B_{r,\mathrm{in}}=B_{r,\mathrm{ph}}$ (see Table \ref{tb_parameter}), and the radial velocity is set to $v_{r,\mathrm{in}}=0$. 
Here, the subscript ``in'' denotes values at the inner boundary.

At the inner boundary $r = r_\mathrm{in}$, a free boundary condition is applied to the inward Els\"asser variables:
\begin{equation}
\label{eq_bc1}
\left.\frac{\partial\bm{z}^{-}_{\perp}}{\partial r}\right|_{\mathrm{in}} = 0.
\end{equation}
The amplitude of the outward Els\"asser variables $\bm{z}_{\perp,\mathrm{in}}^{+}$ is adjusted so that the time-averaged Alfv\'en-wave energy flux $\overline{F_A}_\mathrm{,in}$ remains constant \citep{Shoda2020}, where
\begin{equation}
\label{eq_bc3}
F_{A} = \left(\frac{1}{2}\rho\bm{v}_{\perp}^{2} + \frac{\bm{B}_{\perp}^{2}}{4\pi}\right)v_r - \frac{B_r}{4\pi}(\bm{v}_{\perp}\cdot\bm{B}_{\perp}).
\end{equation}
Here, $\overline{X}$ denotes the time average of $X$ over $2/f_{\mathrm{min}}$, with $f_{\mathrm{min}}$ defined below.
The values of $\overline{F_A}_\mathrm{,in}$ for each run are listed in Tables \ref{tb_parameter2} and \ref{tb_parameter3}.

The spectrum of $\bm{z}_{\perp,\mathrm{in}}^{+}$ is prescribed as
\begin{equation}
\label{eq_bc4}
z_{i,\mathrm{in}}^{+} \propto \sum_{n=0}^{N} \frac{\sin(2\pi f_n t + \phi_{n,i})}{f_n^{5/6}},
\end{equation}
\begin{equation}
\label{eq_bc5}
f_n = \frac{(N-n)f_{\mathrm{min}} + n f_{\mathrm{max}}}{N},
\end{equation}
where $i \in \{x, y\}$, $\phi_{n,i}$ is a random phase, and $N+1$ is the total number of modes, with $N=100$.
The frequency range is limited to $[f_{\min}, f_{\max}]$ (Tables \ref{tb_parameter2}, \ref{tb_parameter3}).
This prescription yields outward Els\"asser energy with a Kolmogorov-type spectrum, $({z}_{i,\mathrm{in}}^{+})^{2} \propto f^{-5/3}$, consistent with observations of coronal transverse-wave spectra \citep{Morton2019}. The driven spectrum remains deep within the MHD regime at the point where it is injected, well below the local proton cyclotron frequency even for the M-dwarf runs. The subsequent nonlinear evolution and dissipation of the injected spectrum are governed not by $f_{\max}$ itself but by the phenomenological strong-turbulence closure of Equations~(\ref{eq_turb1})--(\ref{eq_turb3}), whose dissipation timescale $\lambda_\perp/|z_\perp|$ is independent of the driving-spectrum cutoff.

\begin{deluxetable*}{ccccccccccc}
\tablecaption{Parameters of solar-wind simulations. \label{tb_parameter2}}
\tablehead{
\colhead{run} &
\colhead{AWT} &
\colhead{$B_{\mathrm{cor}}$} &
\colhead{$\overline{F_A}_{\mathrm{,in}}$} &
\colhead{$f_{\min}$} &
\colhead{$f_{\max}$} &
\colhead{$r_\mathrm{cnd}$} &
\colhead{$r_{\mathrm{out}}$} &
\colhead{$f_\mathrm{ph}^\mathrm{open}$} &
\colhead{$t_\mathrm{start}$} &
\colhead{$t_\mathrm{end}$} \\[-0.3em]
\colhead{} &
\colhead{} &
\colhead{[G]} &
\colhead{[$\mathrm{erg\ cm^{-2}\ s^{-1}}$]} &
\colhead{[mHz]} &
\colhead{[mHz]} &
\colhead{[$r_\ast$]} &
\colhead{[$r_\ast$]} &
\colhead{} &
\colhead{[min]} &
\colhead{[min]}
}
\startdata
S1 & included & 13.4 & $1.6\times 10^9$ & 1 & 10 & 10 & 32 & $1.00\times10^{-3}$ & 4000 & 5000 \\
S2 & not included & 13.4 & $0.8\times 10^9$ & 1 & 10 & 10 & 32 & $1.00\times10^{-3}$ & 7000 & 8000 \\
\enddata
\tablecomments{The second column indicates whether AWT is included in the simulation.
$B_{\mathrm{cor}}$ is the coronal magnetic-field strength, and
$\overline{F_A}_\mathrm{,in}$ is the energy flux injected at the inner boundary.
$f_{\min}$ and $f_{\max}$ are the minimum and maximum frequencies of the injected spectrum.
$r_{\mathrm{cnd}}$ is the location where thermal conduction is quenched, and
$r_{\mathrm{out}}$ is the outer boundary.
$f_\mathrm{ph}^\mathrm{open}$ is the photospheric filling factor, and
$t_\mathrm{start}$ and $t_\mathrm{end}$ are the averaging interval.}
\end{deluxetable*}

\begin{deluxetable*}{ccccccccccc}
\tablecaption{Parameters of M-dwarf wind simulations. \label{tb_parameter3}}
\tablehead{
\colhead{run} & \colhead{AWT} & \colhead{$B_{\mathrm{cor}}$} & \colhead{$\overline{F_A}_{\mathrm{,in}}$} & \colhead{$f_{\min}$} & \colhead{$f_{\max}$} & \colhead{$r_\mathrm{cnd}$} & \colhead{$r_{\mathrm{out}}$} & \colhead{$f_\mathrm{ph}^\mathrm{open}$} & \colhead{$t_\mathrm{start}$} & \colhead{$t_\mathrm{end}$} \\[-0.5em]
\colhead{}    & \colhead{}     & \colhead{[G]} & \colhead{[$\mathrm{erg}\ \mathrm{cm}^{-2}\ \mathrm{s}^{-1}$]} &
\colhead{[mHz]} & \colhead{[mHz]} & \colhead{[$r_\ast$]} & \colhead{[$r_\ast$]} & \colhead{} & \colhead{[min]} & \colhead{[min]}
}
\startdata
M1  & included & 10   & $1.76\times10^{9}$ & 4 & 400 & 2 & 176 & $7.19\times10^{-5}$ & 3000 & 4000 \\
M2  & included & 7    & $1.76\times10^{9}$ & 4 & 400 & 2 & 100 & $5.04\times10^{-5}$ & 3000 & 4000 \\
M3  & included & 5    & $1.76\times10^{9}$ & 4 & 400 & 2 & 100 & $3.60\times10^{-5}$ & 3000 & 4000 \\
M4  & included & 3    & $1.76\times10^{9}$ & 4 & 400 & 2 & 100 & $2.16\times10^{-5}$ & 3000 & 4000 \\
M5  & included & 2    & $1.76\times10^{9}$ & 4 & 400 & 2 & 100 & $1.44\times10^{-5}$ & 3000 & 4000 \\
M6  & included & 1    & $1.76\times10^{9}$ & 4 & 400 & 2 & 100 & $7.19\times10^{-6}$ & 3000 & 4000 \\
M7  & included & 0.7  & $1.76\times10^{9}$ & 4 & 400 & 2 & 100 & $5.04\times10^{-6}$ & 7000 & 8000 \\
M8  & included & 0.5  & $1.76\times10^{9}$ & 4 & 400 & 2 & 100 & $3.60\times10^{-6}$ & 5000 & 6000 \\
\hline
M9  & not included  & 10   & $8.78\times10^{8}$ & 4 & 400 & 2 & 176 & $7.19\times10^{-5}$ & 8000 & 10000 \\
M10 & not included  & 7    & $8.78\times10^{8}$ & 4 & 400 & 2 & 100 & $5.04\times10^{-5}$ & 4000 & 5000 \\
M11 & not included  & 5    & $8.78\times10^{8}$ & 4 & 400 & 2 & 100 & $3.60\times10^{-5}$ & 3000 & 4000 \\
M12 & not included  & 2    & $8.78\times10^{8}$ & 4 & 400 & 2 & 100 & $1.44\times10^{-5}$ & 3000 & 4000 \\
M13 & not included  & 1    & $8.78\times10^{8}$ & 4 & 400 & 2 & 100 & $7.19\times10^{-6}$ & 7000 & 8000 \\
M14 & not included  & 0.5  & $8.78\times10^{8}$ & 4 & 400 & 2 & 100 & $3.60\times10^{-6}$ & 3000 & 4000 \\
\enddata
\tablecomments{Columns are the same as Table \ref{tb_parameter2}.}
\end{deluxetable*}

\subsection{Initial condition}\label{sec_model6}

For the solar-wind simulations, the initial conditions are given by  
\begin{equation}
\label{eq_ic1}
    \frac{\rho(r)}{\rho_\mathrm{ph}} = \max\!\left[ \exp\!\left( -\frac{r-r_\ast}{H_\mathrm{ph}}\right),\ 10^{-10}\!\left(\frac{r}{r_\ast}\right)^{-2.5}\right],
\end{equation}
\begin{equation}
\label{eq_ic2}
T(r) = T_{\mathrm{ph}},
\end{equation}
\begin{equation}
\label{eq_ic3}
\bm{v} = 0,\quad \bm{B}_\perp = 0.
\end{equation}
For the M-dwarf simulations, runs M6 and M13 (Table \ref{tb_parameter3}) adopt the same initial conditions as above. 
For all other runs, the solutions from M6 and M13 are used as initial states to reduce the computational cost.

\subsection{Numerical scheme}\label{sec_model7}

For the spatial discretization, we employ a non-uniform radial grid defined as  
\begin{equation}
\label{eq_grid}
    \frac{\Delta r^i}{\Delta r_{\mathrm{min}}} =
    \max \!\left[
        1,\ 
        \min \!\left[
            \frac{\Delta r_{\mathrm{max}}}{\Delta r_{\mathrm{min}}},\ 
            \frac{2\epsilon}{2+\epsilon}\frac{r^{i-1}-r_{\mathrm{exp}}}{\Delta r_{\mathrm{min}}}+1
        \right]
    \right],
\end{equation}
where $r^i$ and $\Delta r^i$ denote the radial position of the $i$th cell center and the cell size, respectively. The expansion of the grid begins beyond $r_{\mathrm{exp}}$, with the ratio of neighboring cell sizes in the expansion region specified by $\epsilon$. We set $r_{\mathrm{exp}} = 1.03\,r_\ast$ and $\epsilon = 0.01$. The minimum grid size $\Delta r_{\mathrm{min}}$ is 20 km for the solar cases and 2 km for the M-dwarf cases, both of which are much smaller than the local pressure scale height. The maximum grid size $\Delta r_{\mathrm{max}}$ is set to 3000 km.

We solve Equations (\ref{eq_basic1})--(\ref{eq_basic7}) using the HLLD Riemann solver \citep{Miyoshi2005} combined with third-order strong-stability-preserving (SSP) Runge--Kutta time integration \citep{Shu1988, Gottlieb2001}. The basic equations are formulated in terms of cross-section--weighted variables \citep{shoda2021corona}. To ensure high spatial accuracy within $r<r_{\mathrm{exp}}$, we employ the fifth-order MP5 reconstruction scheme \citep{Suresh1997}; in the grid-expansion region, the second-order MUSCL reconstruction \citep{Leer1979} is used instead. In regions where $\Delta r = \Delta r_{\mathrm{max}}$, MP5 reconstruction is applied to the transverse velocity and magnetic-field components, while MUSCL reconstruction is used for the remaining variables. Thermal conduction is solved separately using the super-time-stepping method \citep{Meyer2012, Meyer2014}.

\section{Results}\label{sec:results}

Before discussing the simulation results, we outline the physical quantities analyzed. The most fundamental quantities are the time-averaged and root-mean-squared values, each computed by integrating over time from $t_{\rm start}$, when the system reaches a quasi-steady state, to $t_{\rm end}$, the end of the simulation:
\begin{equation}
\label{eq_average}
	\langle X \rangle = \frac{1}{t_{\text{end}} - t_{\text{start}}} \int_{t_{\text{start}}}^{t_{\text{end}}} X \, dt,
\end{equation}
\begin{equation}
\label{eq_rms}
	X_\text{rms} = \sqrt{\langle X^2 \rangle}.
\end{equation}
The values of $t_{\rm start}$ and $t_{\rm end}$ vary depending on the simulation setup. The specific values are listed in Tables \ref{tb_parameter2} and \ref{tb_parameter3}.

The properties of stellar winds are generally discussed in terms of time-averaged quantities. For example, the wind mass-loss rate is defined by the following time averaged value measured at the outer boundary:
\begin{equation}
    \dot{M} = 4\pi r^2f^{\mathrm{open}} \langle \rho v_r \rangle.
\end{equation}
In practice, however, it is not evaluated using the time average of $\rho v_r$, but rather by the product of the time-averaged values of $\rho$ and $v_r$ as follows.
\begin{equation}
    \dot{M} \approx 4\pi r^2f^{\mathrm{open}} \langle \rho \rangle \langle v_r \rangle.
\end{equation}
This approximation replaces $\langle \rho v_r\rangle$ with $\langle\rho\rangle\langle v_r\rangle$, introducing an error equal to the covariance $\mathrm{Cov}(\rho,v_r)\equiv\langle\rho v_r\rangle-\langle\rho\rangle\langle v_r\rangle$. By evaluating this covariance at the outer boundary, we find that the relative error introduced by this approximation is several percent or less.

To investigate the dissipation of Alfv\'en waves, the following Alfv\'en-wave energy conservation law is used as a tool for analysis \citep{Shoda2020, Shimizu2022}.
\begin{align}
    \frac{\partial}{\partial t} e_{\rm A} + \frac{1}{4\pi r^2 f^{\mathrm{open}}} \frac{\partial}{\partial r} L_A = -Q^{\mathrm{wp}} - Q^{\mathrm{turb}}, \label{eq_Qwave1}
\end{align}
where 
\begin{equation}
    e_{\rm A} = \frac{1}{2}\rho \bm{v}_\perp^2 + \frac{\bm{B}_\perp^2}{8\pi},
\end{equation}
\begin{equation}
    L_{\rm A} = 4 \pi r^2 f^{\rm open} F_{\rm A},
\end{equation}
\begin{equation}
    Q^{\mathrm{wp}} = -v_r\frac{\partial}{\partial r}\left(\frac{\bm{B}_\perp^2}{8\pi}\right) + \frac{v_r}{2L} \left(\rho\bm{v}_\perp^2-\frac{\bm{B}_\perp^2}{4\pi}\right),
\end{equation}
\begin{equation}
    L = \left[ \frac{d}{dr} \left( r^2 f^{\rm open} \right) \right]^{-1}.
\end{equation}
We note that $Q^{\rm turb}$ and $F_{\rm A}$ are defined in Equations \eqref{eq_Qturb} and \eqref{eq_bc3}, respectively. $Q^\mathrm{wp}$ represents the work done by wave pressure, describing the conversion of Alfv\'en-wave energy into longitudinal motion through magnetic pressure and centrifugal forces. This term accounts for processes such as nonlinear mode conversion in the chromosphere \citep{Hollweg1982, Kudoh_1999_ApJ}, the parametric decay instability in the solar wind \citep{Reville2018, shoda2018frequency}, and the global wave-pressure acceleration throughout the wind \citep{Alazraki_1971_AandA, Belcher_1971_ApJ, Jacques_1977_ApJ}.

\subsection{Solar wind}\label{sec_result1}
\begin{figure*}[!]
\epsscale{0.9}
  \plotone{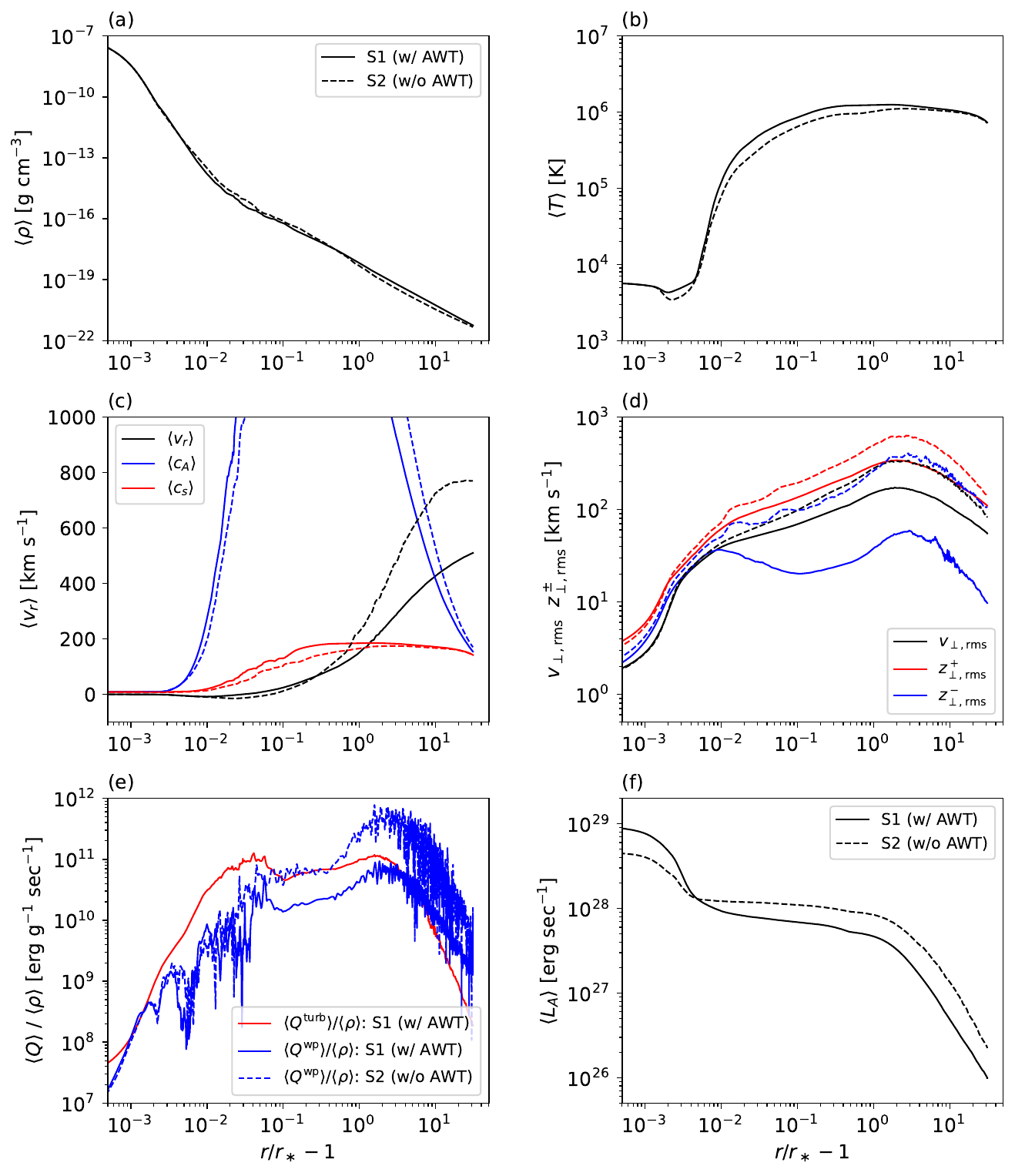}
  \caption{Time-averaged radial profiles for the solar wind simulations (solid lines: Run~S1, dashed lines: Run~S2). Panels show (a) mass density $\langle\rho\rangle$, (b) temperature $\langle T\rangle$, (c) radial velocity $\langle v_r\rangle$, Alfv\'en speed $\langle c_A\rangle$, and sound speed $\langle c_s\rangle$, (d) root-mean-square perpendicular velocity $v_{\perp,\mathrm{rms}}$ and Els\"asser variables $z^\pm_{\perp,\mathrm{rms}}$, (e) wave heating rate $\langle Q^{\mathrm{turb}}\rangle/\langle\rho\rangle$ and work by wave-pressure $\langle Q^{\mathrm{wp}}\rangle/\langle\rho\rangle$, and (f) Alfv\'en-wave luminosity $\langle L_A\rangle$. Since this simulation is for the solar wind, $r_\ast$ is equal to $r_\odot$.}
  \label{fig_Sun1}
\end{figure*}
Because this section discusses the solar wind specifically, we denote the stellar radius $r_\ast$ as $r_\odot$ throughout Section~\ref{sec_result1}. To ensure the consistency with previous studies \citep{suzuki2005making, suzuki2006solar}, we perform a solar-wind simulation without AWT (run S2; dashed lines in Figure \ref{fig_Sun1}). The model reproduces a million-kelvin corona and a solar wind reaching 770 km s$^{-1}$ at the outer boundary ($r_\mathrm{out}=32 r_\odot$). The corresponding mass-loss rate is $\dot{M} = 3.59\times 10^{-14} M_\odot \ \mathrm{yr}^{-1}$, consistent with observations.

The wave-pressure work per unit mass ($\langle Q^\mathrm{wp}\rangle/\langle\rho\rangle$) as a function of radial distance is shown in Figure \ref{fig_Sun1}(e). We note that, for run S2, the turbulent heating rate is zero ($\langle Q^\mathrm{turb}\rangle=0$) due to the absence of AWT. The rise of $\langle Q^{\mathrm{wp}}\rangle/\langle\rho\rangle$ in the lower atmosphere ($r/r_\odot-1<1$) results from nonlinear Alfv\'en-to-slow mode conversion \citep{Hollweg1982, Kudoh_1999_ApJ, suzuki2005making, Arber_2016_ApJ}, which causes atmospheric heating. In the supersonic wind region ($1<r/r_\odot-1$), $\langle Q^{\mathrm{wp}}\rangle/\langle\rho\rangle$ increases further, reflecting plasma acceleration by wave pressure force \citep{Alazraki_1971_AandA, Belcher_1971_ApJ, Jacques_1977_ApJ}.

Solid lines in Figure \ref{fig_Sun1} show the result of Run S1, corresponding to the solar wind simulation with AWT. Like Run S2, this simulation also reproduces a million-kelvin corona, with the wind speed of approximately 500 km s$^{-1}$ at the outer boundary $r_\mathrm{out}=32 r_\odot$, which is well beyond the Alfv\'en point at $r_A\approx10.7\, r_\odot$, where $\langle v_r \rangle = \langle c_A \rangle$. Compared with the case without AWT, the density profile changes little, but the radial velocity decreases. Since solar-wind acceleration arises from energy deposition in the supersonic region, the slower wind speed in Run S1 than in Run S2 suggests that less Alfv\'en-wave energy reached that region. Another notable difference from Run S2 is that the amplitude of the inward Els\"asser variable ($z^-_{\perp,\mathrm{rms}}$) is much smaller than that of the outward one $z^+_{\perp,\mathrm{rms}}$, a natural consequence of dynamical alignment \citep{1980PhRvL..45..144D, Grappin_1982_AandA, Pouquet_1986_PhRvA}. The mass-loss rate ($\dot{M} = 2.89\times 10^{-14} M_\odot \ \mathrm{yr}^{-1}$) is lower than in the case without AWT (Run S2, $3.59\times 10^{-14} M_\odot \ \mathrm{yr}^{-1}$). It is interesting that the mass-loss rate in Run S1 is smaller than in Run S2, despite injecting twice the photospheric energy flux, since the mass-loss rate is expected to scale nearly with the energy input to the corona.

In the chromosphere ($r/r_\odot-1<10^{-2}$), $\langle Q^{\mathrm{turb}}\rangle/\langle\rho\rangle$ exceeds $\langle Q^{\mathrm{wp}}\rangle/\langle\rho\rangle$, indicating that turbulence is the dominant dissipation process of Alfv\'en waves (Figure \ref{fig_Sun1}(e)), consistent with \citet{Shoda2020}. In the low corona ($0.1<r/r_\odot-1<1$), AWT is likewise dominant, implying that the hot corona is maintained by AWT. In contrast, in the supersonic wind region ($r/r_\odot-1>1$), $\langle Q^{\mathrm{wp}}\rangle/\langle\rho\rangle$ becomes comparable to, or exceeds, $\langle Q^{\mathrm{turb}}\rangle/\langle\rho\rangle$, likely owing to the high growth rate of the parametric decay instability in this region \citep{shoda2018frequency}. Nevertheless, this does not mean that turbulence is a minor heating mechanism, since $Q^{\mathrm{wp}}$ here also arises from the ponderomotive acceleration of the solar wind.

Figure \ref{fig_Sun1}(g) shows that the run with AWT exhibits a stronger reduction in the normalized Alfv'en luminosity $\langle L_A\rangle/\overline{L_A}\mathrm{,in}$ within the chromosphere and lower corona ($r/r_\odot-1<10^{-2}$). This indicates that Alfv'en-wave turbulence in the chromosphere causes significant energy loss, underscoring its importance in stellar-wind energetics. As a result of this strong wave dissipation, a hotter chromosphere and corona are produced, as seen in Figure \ref{fig_Sun1}(b))

To assess the realism of the AWT-driven solar wind against recent in-situ observations, we compare two diagnostics of Run S1 with \textit{Parker Solar Probe} (PSP) measurements of the fast solar wind from coronal holes, the regime most directly comparable to our open-field wind model. First, the normalized cross helicity, $\sigma_c \equiv (|z_\perp^+|^2 - |z_\perp^-|^2)/(|z_\perp^+|^2+|z_\perp^-|^2)$, rises from $\sigma_c\approx0.5$ at the photosphere to $\sigma_c\approx0.94$--$0.98$ throughout the corona and wind region within $r<32\,r_\odot$, comparable to, albeit somewhat higher than, the cross helicity of $\sigma_c\approx0.89$ directly measured by PSP in the fast solar wind between $35.5$ and $100\,r_\odot$ during its first orbit \citep{Adhikari_2020_ApJ}; this modest offset is consistent with the expected outward decline of Alfv\'enic imbalance with heliocentric distance. Second, regarding the location of the Alfv\'en point identified above, $r_A\approx10.7\,r_\odot$: direct in-situ crossings of the Alfv\'en critical surface remain rare and inherently local, since the surface is structured rather than spherical. During PSP's eighth encounter, \citet{Kasper_2021_PRL} reported the first sustained sub-Alfv\'enic intervals, spanning heliocentric distances of $16$--$20\,r_\odot$; the longest and most robust of these was traced to a low Mach-number boundary layer associated with a pseudostreamer rather than to open-coronal-hole fast wind, and so does not by itself pin down a single Alfv\'en-surface radius comparable to our result. For the fast-wind regime more directly comparable to our model, \citet{Adhikari_2020_ApJ} used a turbulence-transport model --- calibrated against their own PSP fast-wind measurements at $35$--$100\,r_\odot$ --- to extrapolate an Alfv\'en surface at $\approx11.7\,r_\odot$, in close agreement with our value; we note, however, that this is a model-based estimate rather than a direct observation. Taken together, these comparisons indicate that our AWT model reproduces, at least semi-quantitatively, the physical properties of the fast solar wind established by recent PSP observations. We also considered turbulence-based Solar Orbiter (SolO) studies of the fast wind. \citet{Adhikari_2021_AA} measured the normalized cross helicity separately in fast and slow solar-wind streams using in-situ SolO data, but only for observations obtained after SolO's first perihelion passage, at heliocentric distances of order $110\,r_\odot$ or more --- well outside our simulation domain ($r<32\,r_\odot$) and substantially farther than the PSP measurements used above. \citet{Telloni_2023_AA} used the Metis coronagraph aboard SolO to remotely observe the polar (fast-wind) corona at $1.9$--$5.5\,r_\odot$, inferring the coronal-flow expansion rate and the role of turbulence dissipation in wind acceleration, but did not report the cross helicity or Alfv\'en-point diagnostics used in our comparison. A joint PSP--SolO/Metis analysis by \citet{Adhikari_2022_ApJL} does report $\sigma_c\approx0.96$ and an Alfv\'en surface at $\approx9.22\,r_\odot$, but for wind emerging from the equatorial extension of a coronal hole -- a regime intermediate between fast coronal-hole wind and the slower, streamer-associated wind, rather than the open-coronal-hole fast wind considered here. We therefore adopt the PSP fast-wind measurements of \citet{Adhikari_2020_ApJ} as our primary quantitative benchmark, since they most closely match both the wind regime and the heliocentric-distance range of our simulation.

\subsection{M-dwarf wind}\label{sec_result2}

Having successfully reproduced the solar wind, we now turn to the results of applying our wind model to M-type stars. We focus on a parameter survey of coronal magnetic field strength, aiming to clarify the link between magnetic activity and wind properties. GJ 436 is observationally a low-activity M dwarf: its X-ray luminosity, $L_X=5.7\times10^{26}\ \mathrm{erg\ s^{-1}}$ ($\log L_X=26.76$), measured with \textit{XMM-Newton} \citep{Ehrenreich2015}, is below the mean value for M dwarfs, $\log L_X\approx27.6$. This observed low activity level is consistent with our assumption that rotational effects can be neglected for this target (Section~\ref{chap_model}).
\begin{figure*}[!]
\centering
  \epsscale{0.9}
  \plotone{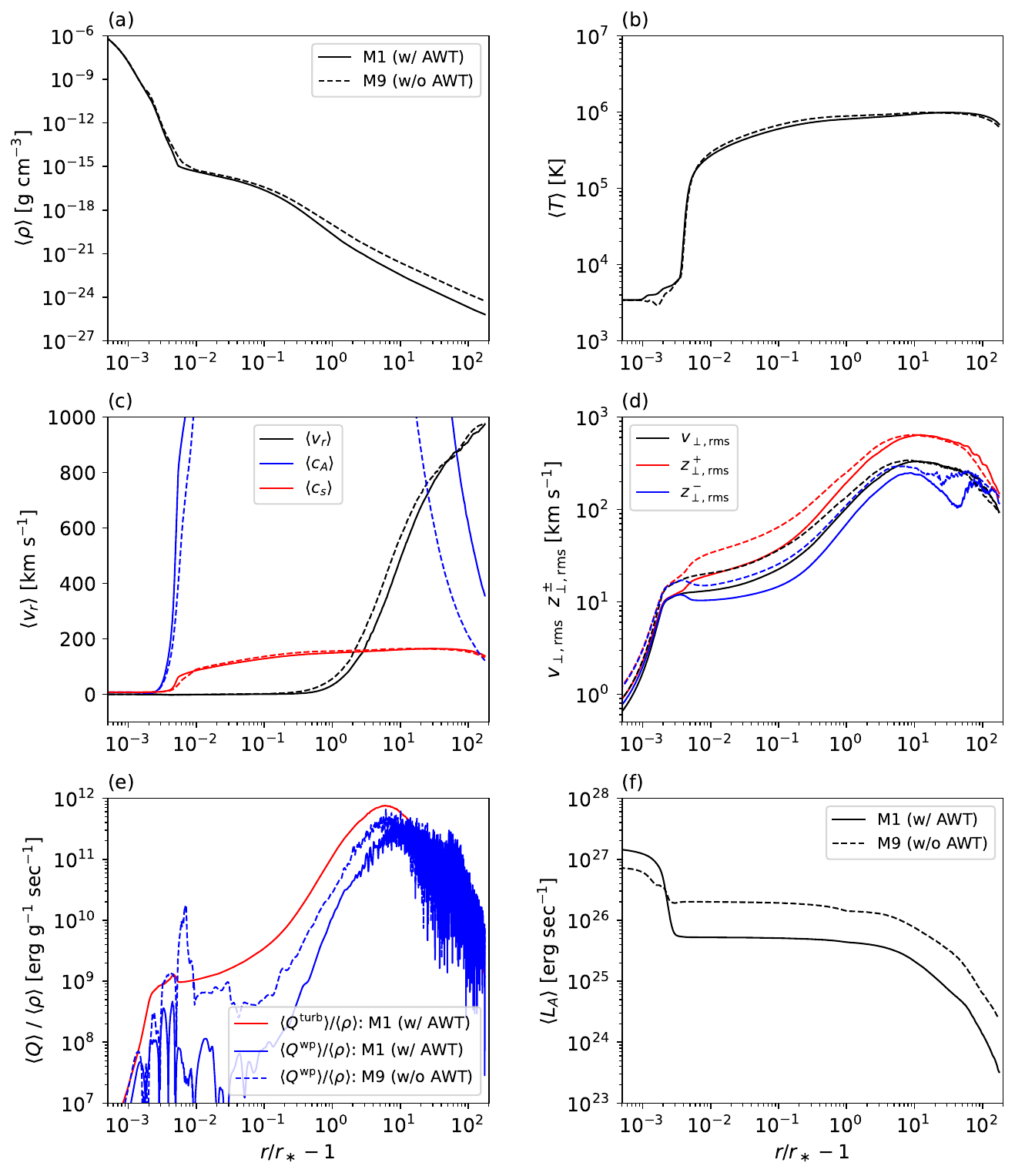}  
  \caption{Time-averaged radial profiles for the M-dwarf wind simulations (solid lines: Run~M1, dashed lines: Run~M9). The panel format is the same as in Figure~\ref{fig_Sun1}.}
  \label{fig_stellar1}
\end{figure*}

Figure \ref{fig_stellar1} presents the results of the M-dwarf simulations (Runs M1 and M9). Both runs reproduce hot coronae exceeding 1 MK and super-Alfv\'enic winds. A major difference, however, arises in the density profile. In Run M1 (solid line), where AWT operates, the wind density is reduced by more than an order of magnitude compared with Run M9. This occurs because, under AWT, most of the injected energy is dissipated in the chromosphere, as shown in Figure \ref{fig_stellar1}(f), where the normalized wave luminosity decreases more strongly in Run M1 than in Run M9. Enhanced wave dissipation also leads to higher chromospheric temperatures in Run M1 than in Run M9 (panel (b)). Consequently, the amplitude of the upward-propagating Els\"asser variable $z^+_{\perp,\mathrm{rms}}$ is smaller (panel (d)), the sonic point shifts outward (panel (c)), and the wind density decreases. The estimated mass-loss rate in Run M1 is $3.93\times 10^{-17} M_\odot \ \mathrm{yr^{-1}}$, far below the observational estimate of $\dot{M}=(0.45\text{-}2.5)\times 10^{-15} M_\odot \ \mathrm{yr^{-1}}$ \citep{Vidotto2017} and the previous estimate of $\dot{M} \sim 10^{-15} M_\odot \ \mathrm{yr^{-1}}$ for an M3.5 star \citep{Sakaue2021}.
\begin{figure*}[!]
  \epsscale{1.0}
  \plotone{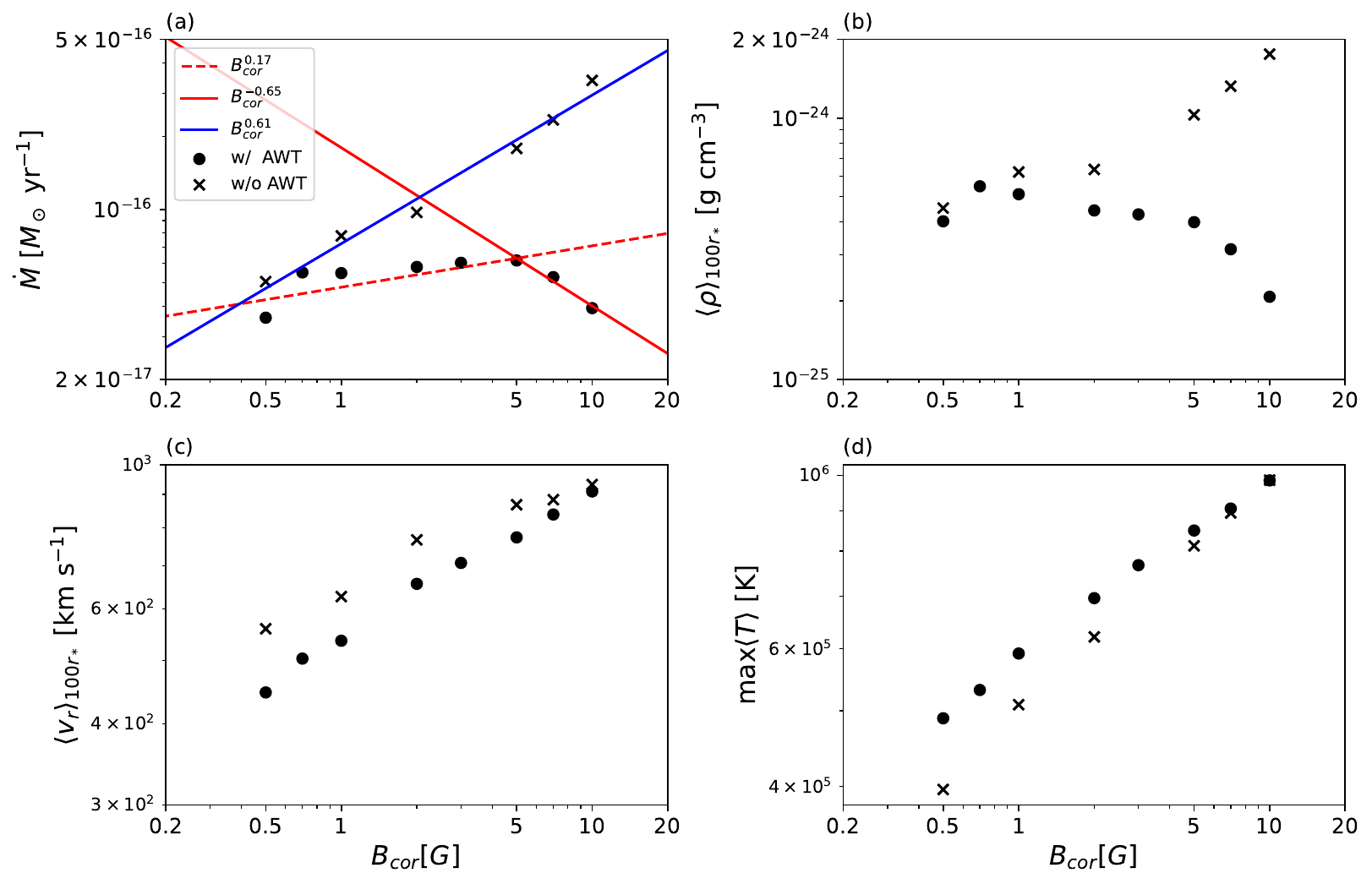}
  \caption{Results of the parameter survey varying $B_{\mathrm{cor}}$ for the M-dwarf wind simulations with AWT (filled circles: Runs~M1--M8) and without AWT (crosses: Runs~M9--M14). Panels show (a) mass-loss rate, (b) density, (c) radial velocity at $r=100\,r_\ast$, and (d) maximum temperature within the simulation domain.}
  \label{fig_Bcor}
\end{figure*}

To examine how the mass-loss rate depends on the coronal magnetic field, we conduct a parameter survey listed in Table \ref{tb_parameter3}. The results are shown in Figure \ref{fig_Bcor}. In the cases without AWT (runs M9--M14, marked with crosses), the mass-loss rate increases monotonically with $B_\mathrm{cor}$, consistent with \citet{Sakaue2021}. In contrast, simulations including AWT (runs M1--M8, marked with filled circles) show a markedly different behavior. The mass-loss rates are lower than those without AWT and reach values one order of magnitude smaller at $B_\mathrm{cor}=10\ \mathrm{G}$. Moreover, the trend differs above and below $B_\mathrm{cor}=5\ \mathrm{G}$: the rate increases slightly for $B_\mathrm{cor}<5\ \mathrm{G}$ but decreases for $B_\mathrm{cor}>5\ \mathrm{G}$. Among other quantities (panels (b)--(d)), only the wind density exhibits a similar trend to the mass-loss rate.
\begin{figure*}[!]
  \epsscale{0.9}
  \plotone{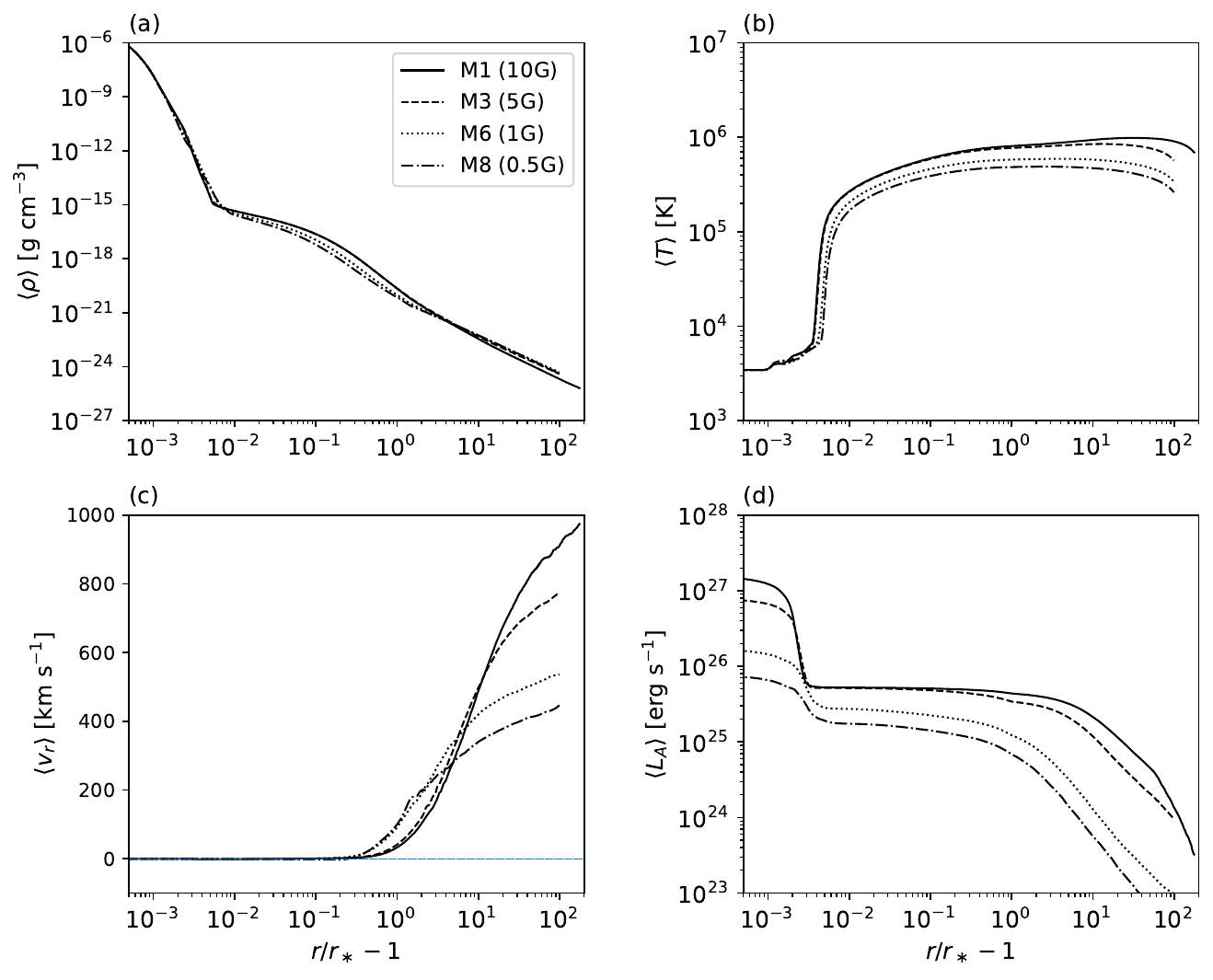}
    \caption{Time-averaged radial profiles from the M-dwarf wind simulations, comparing cases with different $B_{\mathrm{cor}}$ values. Solid, dashed, dotted, and dash-dotted lines correspond to $B_{\mathrm{cor}}=10$~G (Run~M1), $5$~G (Run~M3), $1$~G (Run~M6), and $0.5$~G (Run~M8), respectively. Panels show (a) mass density $\langle\rho\rangle$, (b) temperature $\langle T\rangle$, (c) radial velocity $\langle v_r\rangle$, and (d) Alfv\'en-wave luminosity $\langle L_A\rangle$.
    }
  \label{fig_weakB1}
\end{figure*}

Figure \ref{fig_weakB1} compares the radial profiles of physical variables for Runs M1, M3, M5, and M10. Panel (a) shows that, when the coronal magnetic field is weak ($B_{\mathrm{cor}} \le 5$ G), the density profiles are nearly identical in the chromosphere ($r/r_\ast-1 < 10^{-2}$) and wind region ($r/r_\ast-1 > 1$), but the lower corona becomes denser with larger $B_{\mathrm{cor}}$. With larger $B_{\mathrm{cor}}$, a hotter atmosphere (panel (b)) and faster wind (panel (c)) are produced. The Alfv\'en luminosity $\langle L_A\rangle$ (panel (d)) decreases with increasing $B_{\mathrm{cor}}$, meaning that more Alfv'en wave energy is dissipated in the chromosphere.

The dependence on $B_{\mathrm{cor}}$ (in the weak-field regime) can be interpreted straightforwardly. A larger $B_{\mathrm{cor}}$ increases the Alfv\'en-wave luminosity $\langle L_A\rangle$ at the coronal base (panel (d)) and reduces the correlation length $\lambda_\perp$, thereby enhancing coronal heating. Accordingly, higher $B_{\mathrm{cor}}$ produces higher coronal temperatures and a larger density scale height. In the wind region ($r/r_\ast-1>10$), stronger $B_{\mathrm{cor}}$ also yields greater $\langle L_A\rangle$ (panel (d)), which enhances wave pressure and drives faster winds (panel (c)).

The density profiles are nearly identical for runs M3 and M1 in the stellar atmosphere ($r/r_\ast-1 \lesssim 10$; Figure \ref{fig_weakB1}(a)), indicating that the near-surface density is insensitive to $B_{\mathrm{cor}}$. However, farther out ($r/r_\ast-1 > 10$), the density becomes slightly lower for stronger $B_{\mathrm{cor}}$ (Figure \ref{fig_Bcor}(b)). A similar trend is seen for the temperature and wind speed (Figures \ref{fig_weakB1}(b) and (c)): they remain almost unchanged in the subsonic region, but become higher in the supersonic wind for larger $B_{\mathrm{cor}}$. This behavior suggests that increasing $B_{\mathrm{cor}}$ does not enhance heating in the subsonic region, whereas it leads to more heating and acceleration in the supersonic wind region.

This interpretation is supported by the Alfv'en-wave luminosity profiles (Figure \ref{fig_weakB1}(d)). First, the wave luminosity at the coronal base ($10^{-2}<r/r_\ast-1<10^{-1}$) is nearly identical between $B_{\mathrm{cor}}=5$ G and $10$ G, indicating that the energy input into the corona is essentially unchanged. Second, for $B_{\mathrm{cor}}=10$ G, the wave luminosity decreases more slowly with radius, implying that a larger fraction of wave energy propagates to greater heights before dissipating. Together, these results indicate that the chromospheric and coronal-base heating rates are nearly constant, while the additional energy that survives to larger radii enhances heating and wind acceleration in the supersonic region.

It is also noteworthy that the Alfv'en-wave luminosity at the photosphere is nearly twice as large for $B_{\mathrm{cor}}=10$ G as for $B_{\mathrm{cor}}=5$ G, reflecting the larger flux-tube filling factor at the surface. Nevertheless, the luminosities converge to nearly the same value at the coronal base, implying that more wave energy is dissipated between the photosphere and corona in the stronger-field case. Because the decrease of wave luminosity corresponds to wave dissipation, this result indicates enhanced chromospheric dissipation for larger $B_{\mathrm{cor}}$. This enhanced dissipation can be attributed to two effects: A larger reflection coefficient of Alfv'en waves and a shorter correlation length $\lambda_\perp$ in the chromosphere, both of which lead to stronger turbulent dissipation.

\subsection{Predicted scaling laws of M-dwarf mass-loss rate \label{sec_massloss_scaling_law}}

In our parameter survey, the coronal magnetic field strength $B_{\mathrm{cor}}$ ranges from 0.5 to 10 G, which is smaller than the observed value of $\sim 20$ G for GJ 436 \citep{Cristofari2023}. To estimate the mass-loss rate for $B_{\mathrm{cor}}=20\ \mathrm{G}$, we extrapolate our simulation results using a power-law scaling derived from the strong-field cases with AWT (red solid line in Figure \ref{fig_Bcor}(a)):
\begin{equation}
\label{eq_scaling1}
	\dot{M} = 6.3\times 10^{-17}\ M_\odot\ \mathrm{yr}^{-1}  \left(\frac{B_{\mathrm{cor}}}{5\ \mathrm{G}}\right)^{-0.65}.
\end{equation}
This scaling predicts a mass-loss rate of $\dot{M}=2.6\times 10^{-17}\ M_\odot\ \mathrm{yr}^{-1}$ for $B_{\mathrm{cor}}=20\ \mathrm{G}$, which is more than an order of magnitude smaller than the observational estimate of $\dot{M}=(0.45\text{--}2.5)\times 10^{-15}\ M_\odot\ \mathrm{yr}^{-1}$ \citep{Vidotto2017} and still below the estimate of $\dot{M}=(0.51\text{--}1.1)\times 10^{-15}\ M_\odot\ \mathrm{yr}^{-1}$ for an M3.5 star \citep{Sakaue2021}. We note that the GJ~436 astrospheric mass-loss estimate used above is the same measurement subsequently compiled by \citet{Wood_2021_ApJ} in their systematic Ly$\alpha$-astrosphere survey of M-dwarf winds. That survey finds that the winds of most M dwarfs are weaker than or comparable to the solar wind, with mass loss increasing with X-ray activity only weakly and with substantial star-to-star scatter; GJ~436, being an X-ray-faint, low-activity M dwarf (Section~\ref{sec_result2}), is unexceptional in this context. Our predicted mass-loss rate therefore falls short not only of the single GJ~436 measurement but of the broader trend defined by the observed M-dwarf population, reinforcing that the discrepancy reflects a genuine limitation of the model rather than a peculiarity of the comparison star.

For the runs without AWT, we similarly derive a scaling law (blue solid line in Figure \ref{fig_Bcor}(a)):
\begin{equation}
\label{eq_scaling2}
	\dot{M} = 4.5 \times 10^{-16}\ M_\odot\ \mathrm{yr^{-1}} \left(\frac{B_{\mathrm{cor}}}{20\ \mathrm{G}}\right)^{0.61}.
\end{equation}
At $B_{\mathrm{cor}}=20\ \mathrm{G}$, this relation predicts a mass-loss rate of $\dot{M}=4.5 \times 10^{-16}\ M_\odot\ \mathrm{yr^{-1}}$, which is in good agreement with the estimate by \cite{Sakaue2021} obtained from simulations without AWT.

\section{Discussion} \label{sec:Discussion}  

\subsection{Comparison with previous studies}\label{sec_discussion_compare}

Previous one-dimensional MHD simulations have shown that heating of the solar wind is predominantly driven by Alfv'en-wave turbulence (AWT) \citep{shoda2018self}. Nevertheless, \citet{suzuki2005making, suzuki2006solar} demonstrated that the solar wind can still be reproduced in the absence of AWT. Our solar runs without AWT confirm this result: the wind is sustained primarily through shock heating. By contrast, when AWT is included, the coronal heating is dominated by turbulent dissipation, consistent with the findings of \citet{shoda2018self}. These results indicate that, even without AWT, shock dissipation provides sufficient coronal heating to maintain a quasi-steady solar wind.

In contrast to \citet{shoda2018self}, our simulations explicitly include turbulent dissipation in the chromosphere. This chromospheric AWT limits the Alfv'en-wave Poynting flux entering the corona, thereby reducing coronal heating. Consequently, in contrast to the parameter survey results of \citet{shoda2018self}, the presence (or efficiency) of chromospheric turbulence has a strong impact on the mass-loss rate. Considering that the turbulent dissipation timescale \citep[$\sim 1000\ \mathrm{s}$,][]{Cranmer2007} is shorter than the chromospheric Alfv'en crossing time ($\sim 100$--$300\ \mathrm{s}$), our results imply that even modest wave dissipation in the chromosphere can substantially influence the solar-wind mass flux \citep{Matsuoka_2024_ApJ}. This has important implications for global solar-wind models: most existing models set their lower boundary near the top of the chromosphere, but future models with photospheric lower boundaries will need to account for chromospheric turbulence and wave dissipation to reproduce realistic mass-loss rates.

\citet{Sakaue2021} performed stellar-wind simulations for a range of M dwarfs and found mass-loss rates comparable to those from our simulations without AWT, but systematically larger than those from our simulations with AWT (Section \ref{sec_massloss_scaling_law}). Their setup differs from ours in several respects, including the target stellar parameters and the coordinate system. They also conducted parameter surveys by varying the coronal magnetic-field strength and the energy flux injected from the stellar surface. However, their variation of magnetic-field strength was implemented by changing the flux-tube filling factor, whereas our parameter survey varied the filling factor of the flux tube. In terms of the energy-flux survey, our approach is consistent with theirs in that both effectively change the photospheric Alfv\'en luminosity $\langle L_A\rangle$. In their simulations, the mass-loss rate increases with the injected energy flux, consistent with our AWT-free runs. By contrast, our runs with AWT show the opposite trend for strong coronal fields, highlighting that turbulent dissipation in the chromosphere fundamentally alters how the mass-loss rate responds to energy input.

\subsection{Possible Causes of the Low Mass-Loss Rate of M-dwarf Winds}\label{sec_discussion2}

Because no in-situ measurements of the M-dwarf wind plasma are available, we cannot validate the M-dwarf AWT model as directly as we validated the solar-wind model against PSP observations in Section~\ref{sec_result1}. That solar validation, however, indicates that the underlying AWT mechanism itself reproduces the physical properties of an observed stellar wind reasonably well; this lends support to the interpretation, developed below, that the low mass-loss rate obtained for GJ~436 reflects uncertainty in the chromospheric turbulence parameters assumed for M dwarfs rather than a fundamental flaw in the AWT framework. Several factors may account for the discrepancy between our AWT simulations and observational estimates of the mass-loss rate. One important factor is the uncertainty in the input parameters. The injected Alfv\'en wave flux $\overline{F_A}_\mathrm{,in}$ plays a critical role in determining the mass-loss rate of M-dwarf winds. In our simulations, $\overline{F_A}_\mathrm{,in}$ was scaled from solar values, which may not accurately represent the actual stellar conditions. In addition, \citet{Shulyak2019} showed that the filling factor and magnetic-field distribution vary from star to star, introducing further uncertainty into the estimation of the Alfv\'en-wave luminosity.

In addition to the uncertainties in the input parameters discussed above, our M dwarf wind model may omit several physical processes that could contribute to wind acceleration and heating. Notable examples include stellar rotation and magnetic reconnection with surrounding closed magnetic field lines. For the solar wind, approximately half of the total energy flux is supposed to be supplied through the transfer of energy from surrounding closed magnetic field regions \citep{Iijima2023}. If the efficient transfer of energy from closed to open field regions operates similarly in M dwarfs, this mechanism could help reconcile our low predicted mass loss rates with the higher values inferred from observations. These considerations highlight the need for future studies to address both the uncertainties in the input parameters and the influence of potentially missing physical processes. The effects of stellar rotation and reconnection can be incorporated into one-dimensional simulations through appropriate boundary conditions or additional phenomenological source terms \citep{Shoda2020, Shoda2023}. Moreover, three-dimensional simulations will likely be indispensable for capturing inherently multidimensional processes, such as turbulence, that cannot be faithfully represented in one-dimensional models \citep{van_Ballegooijen_2016_ApJ, Chandran_2019_JPlPh}. Finally, given the cooler atmospheric conditions of M dwarfs, it may be necessary to account for differences in radiative processes relative to the solar atmosphere, as well as for ambipolar diffusion \citep{Matsuoka_2024_ApJ, Suzuki_2025_PASJ}.

The smaller mass-loss rate in our model is attributed to AWT in the chromosphere. Our model may overestimate the Alfv\'en-wave dissipation in M dwarfs, including GJ 436. One possible explanation is that the turbulent correlation length in M dwarfs is significantly larger than the value assumed in our model. An alternative possibility is that turbulence does not fully develop in the chromospheres of M dwarfs at all. This possibility is supported by the concurrent study of \citet{Shoda_2026_MNRAS}, who tested the effect of suppressing chromospheric turbulent dissipation across a broad ensemble of solar magnetic flux tubes spanning coronal field strengths comparable to, and substantially exceeding, those considered here. Their suppression is motivated by radiation-MHD simulations indicating that chromospheric Alfv\'en waves are driven by flux-tube rotation and dissipate preferentially at flux-tube boundaries rather than through an interior turbulent cascade. Suppressing chromospheric dissipation increased the coronal mass flux by up to an order of magnitude and brought their models into agreement with the empirical relation between coronal magnetic-field strength and wind mass flux, whereas models retaining the standard turbulence treatment --- essentially the same treatment adopted here --- systematically underproduced the mass flux at strong fields, mirroring the discrepancy we find for GJ 436. Recent studies on solar coronal heating have drawn attention to energy transport by photospheric swirl motions \citep{Yadav_2020_ApJ, Breu_2022_AA, Finley_2022_AA}, often referred to as magnetic tornadoes \citep{Wedemeyer-Bohm_2012_Natur, Kuniyoshi_2023_ApJ}. Numerical simulations suggest that the onset of magnetic tornadoes steepens the magnetic energy spectrum \citep{Kuniyoshi_2023_ApJ}, implying a suppression of the turbulent cascade. If the energy transport from the photosphere to the corona in M dwarfs is similarly dominated by magnetic tornadoes, the level of turbulent dissipation in the chromosphere could be much smaller than assumed in our model, regardless of the correlation length. Future numerical simulations of stellar winds should therefore carefully examine the modeling of the chromospheric turbulence.

\section{Summary and Conclusion}\label{chap:Conclusion}

We performed one-dimensional MHD simulations of solar and M-dwarf winds including radiative cooling, thermal conduction, gravity, and phenomenological Alfv\'en-wave turbulence (AWT). To assess the role of AWT in the solar wind, we compared simulations with and without AWT. In the AWT cases, the wind was sustained primarily by turbulent heating, whereas in the absence of AWT it was maintained by shock dissipation. Without AWT, about $30\%$ of Alfv\'en waves were reflected and propagated inward, while with AWT, dynamic alignment suppressed reflection and produced an almost pure outward-propagating Alfv\'enic state. Because the latter agrees better with observations, the model including AWT provides a more realistic description of the solar wind.

For the M-dwarf winds, we performed simulations targeting GJ 436, which has a relatively weak magnetic field, to investigate whether turbulence or shocks play the dominant role in wind driving. We found that turbulent heating is significant in the corona, indicating that AWT is the primary driver of the wind in GJ 436. In addition, we carried out a parameter survey of the coronal magnetic field strength both with and without AWT to examine the dependence of the mass-loss rate. In our model, the energy flux injected at the stellar surface varies with $B_{\mathrm{cor}}$. In the presence of AWT, the behavior of the mass-loss rate changes around $B_{\mathrm{cor}}=5\ \mathrm{G}$. For $B_{\mathrm{cor}}<5\ \mathrm{G}$, the mass-loss rate increases with $B_{\mathrm{cor}}$ because the energy flux entering the corona becomes larger. Without AWT, a larger fraction of the Alfv\'en-wave energy reaches the supersonic region without being dissipated, leading to a decrease in the mass-loss rate for $B_{\mathrm{cor}}>5\ \mathrm{G}$. The mass-loss rate predicted for $B_{\mathrm{cor}}=20\ \mathrm{G}$, which is a realistic magnetic-field strength for M dwarfs, is significantly smaller than both observational estimates and previous simulation results.

Without AWT, the mass-loss rate increases monotonically with $B_{\mathrm{cor}}$. The primary difference between the two cases lies in the dissipation of Alfv\'en waves in the chromosphere. In the presence of AWT, turbulent dissipation in the chromosphere limits the Alfv\'en-wave flux reaching the corona, whereas without AWT, variations in the surface flux are directly reflected in the coronal flux. At realistic magnetic-field strengths, the mass-loss rate obtained without AWT is larger than that with AWT and is closer to the values inferred from observations and other simulations.

Our results indicate that the turbulent dissipation in the chromosphere may be overestimated in our M-dwarf wind model. Future simulations of stellar winds should therefore carefully constrain the parameters governing the efficiency of turbulent dissipation in the chromosphere.

\begin{acknowledgments}
T.Y. is supported by JSPS KAKENHI Grant Number JP21H04492.
M.S. is supported by JSPS KAKENHI Grant Number JP24K00688.
This manuscript was prepared with the assistance of Claude (Anthropic), a large language model, which was used to help draft and revise text in response to the referee's and editor's comments. All AI-assisted content was reviewed and verified by the authors, who take full responsibility for the accuracy of the manuscript.
\end{acknowledgments}

\begin{contribution}
K.S. carried out the numerical simulations and analyzed the data. T.Y. secured the computational resources for the simulations and supervised the overall project. M.S. provided the numerical simulation code and proposed the data analysis methods. All authors contributed to the discussion of the results and to the writing of the manuscript.
\end{contribution}

\software{matplotlib \citep{Hunter_2007_CSE}, NumPy \citep{Harris_2020_Nature}, Claude (Anthropic)}

\bibliography{reference}
\bibliographystyle{aasjournal}
\end{document}